\documentclass{ws-ijmpb}

\newcommand{\bq}{{\bf q}}
\newcommand{\bk}{{\bf k}}

\newcommand{\nubar}{\bar{\nu}}
\newcommand{\rhobar}{\bar{\rho}}

\newcommand{\Hhat}{\hat{H}}
\newcommand{\stilde}{\tilde{s}}

\begin{document}

\markboth{M.\ O.\ Goerbig}
{Second Generation of Composite Fermions and the Self-Similarity of the FQHE}

\catchline{}{}{}{}{}

\title{Second Generation of Composite Fermions and the Self-Similarity of 
the Fractional Quantum Hall Effect}

\author{\underline{M.\ O.\ Goerbig}$^{1,2,}$\footnote{E-mail: goerbig@lps.u-psud.fr}, P.\ Lederer$^2$, and C.\ Morais Smith$^{1,3}$}
\address{$^1$D\'epartement de Physique, Universit\'e de Fribourg, P\'erolles,
CH-1700 Fribourg, Switzerland.\\
$^2$Laboratoire de Physique des Solides, B\^at. 510 (associ\'e au CNRS),
91405 Orsay cedex, France.\\
$^3$Institute for Theoretical Physics, Utrecht University, Leuvenlaan 4,\\
3584 CE Utrecht, The Netherlands.}

\maketitle

\begin{history}
\received{25 June 2004}
\revised{20 August 2004}
\end{history}

\begin{abstract}
A recently developed model of interacting composite 
fermions, is used to investigate different composite-fermion phases. Their
interaction potential allows for the formation of both solid and new 
quantum-liquid phases, which are interpreted in terms of second-generation
composite fermions and which may be responsible for the fractional quantum
Hall states observed at unusual filling factors, such as $\nu=4/11$. Projection
of the composite-fermion dynamics to a single level, involved in the 
derivation of the Hamiltonian of interacting composite fermions, 
reveals the underlying self-similarity of the model.
\end{abstract}


\section{Introduction}

Most fractional quantum Hall effects (FQHE) may be understood as an 
integer quantum Hall effect (IQHE) 
of a quasi-particle, a so-called composite fermion (CF),
which consists of an electron and a vortex-like object with vorticity 
$2s$.\cite{jain}
The formation of CFs is due to the presence of strong correlations between
the interacting electrons in a partially filled Landau level (LL), the filling 
of which is characterized by the ratio $\nu=n_{el}/n_B$ between the electronic
and the flux densities, $n_{el}$ and $n_B=B/(h/e)$, respectively.
Because of its fractional charge,
the CF experiences a reduced coupling $(eB)^*=eB/(2sp+1)$ to the external 
magnetic field $B$ and, in an approximate sense, forms LLs (CF-LLs) 
itself.\cite{MS}
The FQHE arises when the CF filling factor $\nu^*=hn_{el}/(eB)^*$ is an 
integer $p$. 
Because the electronic and CF filling factors are 
related by $\nu=\nu^*/(2s\nu^*+1)$,
this type of FQHE is expected for the series $\nu=p/(2sp+1)$.

Recently, Pan {\it et al.} have observed a FQHE at $\nu=4/11$, which 
may not be described as an IQHE of CFs, but
should rather be viewed as a FQHE of CFs at $\nu^*=1+1/3$.\cite{pan} This
discovery has renewed the interest in a possible self-similarity of the
FQHE, investigated by Mani and v. Klitzing on the basis of scaling 
transformations.\cite{mani1} It is natural to interpret such new FQHE states
in terms of higher-generation CFs: the $4/11$ state may thus
be viewed as an IQHE of second-generation CFs 
(C$^2$Fs).\cite{smet,goerbig04,lopezfradkin} However, recent numerical 
investigations by W\'ojs {\it et al.} indicate that the interaction potential 
between quasi-particles in $1/3<\nu<2/5$ is not sufficiently short-range to 
allow for the formation of higher-generation CFs.\cite{wojs} Furtheremore,
numerical-diagonalization studies in the CF basis reveal an alternation between
compressible and incompressible states at $\nu=4/11$ with varying CF 
number and have not provided a conclusive answer about the nature of this state
in the thermodynamic limit.\cite{jainbla}

Here, we investigate the self-similarity of the FQHE within a model of
interacting spin-polarized
composite fermions at $\nu^*\neq p$, which we have recently derived
in the framework of Murthy and Shankar's Hamiltonian theory.\cite{MS}
The self-similarity is found in the mathematical structure of the model: its
Hamiltonian has the same form as that of electrons restricted to a single LL
if one replaces the electronic by the CF interaction potential. Furthermore, 
the CF 
density operators, restricted to a single CF-LL, satisfy the same commutation
relations as the projected operators of the electron density, in terms of a
renormalized magnetic length $l_B^*=\sqrt{\hbar/(eB)^*}$.\cite{goerbig04}
However, the CF interaction potential, which has been derived within the model,
and that for electrons in a single LL have a {\it different} form and cannot
be related to each other by simple rescaling of the magnetic length. The 
existence of higher-generation CFs is therefore not guaranteed by the 
self-similarity of the model. Nevertheless, detailed energy calculations for 
competing CF quantum-liquid and solid phases, such as bubbles and stripes of 
CFs, have been performed within the model and indicate the stability of some 
C$^2$F states.\cite{goerbig08} 

\section{Model of Interacting Composite Fermions}

At $p<\nu^*<p+1$, one is confronted with a ground-state degeneracy in
a model of non-interacting CFs. This degeneracy is lifted by the residual CF
interactions, which may be taken into account in the 
Hamiltonian\cite{goerbig04} ($l_B=\sqrt{\hbar/eB}\equiv 1$)
\begin{equation}\label{equ01}
\Hhat(s,p)=\frac{1}{2A}\sum_{\bq}v_{s,p}^{CF}(q)\rhobar^{CF}(-\bq)
\rhobar^{CF}(\bq),
\end{equation}
with the CF-interaction potential, given in terms of 
Laguerre polynomials $L_p(x)$,
\begin{equation}\label{equ02}
v_{s,p}^{CF}(q)=\frac{2\pi e^2}{\epsilon(q) q}e^{-q^2l_B^{*2}/2}
\left[L_p\left(\frac{q^2l_B^{*2}c^2}{2}\right)-c^2e^{-q^2/2c^2}
L_p\left(\frac{q^2l_B^{*2}}{2c^2}\right)\right]^2
\end{equation}
and the vortex charge $c^2=2ps/(2ps+1)$.\cite{goerbig04} This model
describes low-energy excitations within the same CF-LL, and the restricted
CF density operators $\rhobar^{CF}(-\bq)$ satisfy the commutation relations
$[\rhobar^{CF}(\bq),\rhobar^{CF}(\bk)]=2i\sin\left[(\bq\times\bk)_zl_B^{*2}/
2\right]\rhobar^{CF}(\bq+\bk)$. In contrast to prior investigations,
where the interaction potential has been constructed from a
few numerically determined pseudopotentials,\cite{leescjain,wojs} here,
it has been derived in a set of transformations,\cite{goerbig04} in the
framework of the Hamiltonian theory of the FQHE.\cite{MS} Inter-CF-LL 
excitations are taken into account with the help of a $q$-dependent dielectric 
function $\epsilon(q)$ in the CF interaction potential 
(\ref{equ02}).\cite{goerbig04}

Because of its similarity with the electronic model, the Hamiltonian 
(\ref{equ01}) may be analyzed by the same techniques as the one
of electrons restricted to a single level.\cite{goerbig08} The energy
$E_{coh}^L(s,p;\stilde)$ 
of the quantum-liquid phases at $\nubar^*=1/(2\stilde+1)$, where $\stilde$ is
an integer, and $\nubar^*=\nu^*-p$ is the partial filling of the $p$-th CF-LL,
may be calculated in Laughlin's wave function approach.\cite{laughlin} 
Note that we restrict the discussion to Laughlin liquids here although other 
incompressible liquids may also play a role at certain CF filling factors. At
$\nubar^*\neq 1/(2\stilde+1)$, the energy $\Delta_{s,p}^{qp/qh}(\stilde)$
of the excited quasiparticles 
[for $\nubar^*>1/(2\stilde+1)$] or quasiholes [for $\nubar^*<1/(2\stilde+1)$]
has to be taken into account. The latter may be interpreted as C$^2$Fs or 
C$^2$F-holes excited to a higher level, which raise the energy of 
the quantum-liquid phases away from $\nubar^*=1/(2\stilde+1)$, where local
minima in form of cusps are obtained. They are at the origin of the 
incompressibility of the quantum liquids.
The energy of CF-solid phases may be calculated in the Hartree-Fock 
approximation, which has been used in the study of electron-solid phases
in higher LLs.\cite{FKS,moessner,goerbig05} 
The expressions for the energies of the quantum-liquid and CF solid phases may 
be found in Ref.~\refcite{goerbig08}.

\section{Results}

The results for the energies of the different CF phases are shown in the 
Fig.1. In contrast to Ref.~\refcite{goerbig08}, the energies are calculated 
with a screened interaction potential.
The quantum-liquid (C$^2$F) phases are stable around
$\nu^*=1+1/3$ and $1+1/5$, which correspond to the electronic fillings 
$\nu=4/11$ and $6/17$, respectively. Whereas a spin-polarized FQHE at 
$\nu=4/11$ has been observed by Pan {\it et al.}, only a tiny local 
minimum in the longitudinal resistance hints to the existence of a possible 
$6/17$ state.\cite{pan} Indeed, a C$^2$F state is close in energy to
a CF Wigner crystal ($M=1$), the energy of which is lowered if one takes into
account an underlying impurity potential: a solid phase may take advantage of
the minima of such a potential by deformation of its crystalline structure, in
analogy with the electronic case.\cite{goerbig05} It is therefore not clear 
whether the quantum liquid at $\nu=6/17$ survives in the presence of 
impurities. At half-filling, one finds a CF stripe phase, which may give rise
to an anisotropic longitudinal resistance, which has recently been observed at
$\nu=13/8$.\cite{fischer} While these results are similar to those 
obtained for the competing {\it electronic} phases in the first excited 
LL ($n=1$),\cite{goerbig05} as one may expect from self-similarity arguments,
there are differences due to the different form of the CF
interaction potential. In $p=1$, the $1/3$ state, {\it e.g.}, is more stable 
than the $1/5$ state in the CF model, and a two-CF bubble phase  ($M=1$) is 
unstable (see figure), in contrast to the $n=1$-LL.\cite{goerbig05} 

\begin{figure}[htbp]
\centerline{\psfig{file=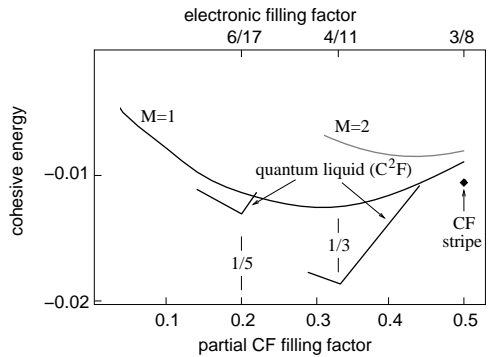,width=6.5cm}}
\vspace*{8pt}
\label{fig01}
\caption{Cohesive energies of the different CF phases, in units of
$e^2/\epsilon l_B$.}
\end{figure}

Note that the CF Wigner crystal has a lower energy than the quantum liquid 
above $\nu\simeq 0.44$. This transition point is shifted to lower densities 
if one takes into account impurity effects and
the repulsive interaction between the excited
C$^2$Fs, which leads to non-linear slopes in the quantum-liquid energies at 
$\nubar^*\neq1/(2\stilde+1)$.
Impurities favor the CF Wigner crystal also
at small values of $\nubar^*$. Therefore, the C$^2$F phases
are surrounded by insulating CF Wigner crystals, and this may lead
to a {\it reentrance phenomenon of the FQHE},\cite{goerbig08}
in analogy with the reentrance
of the IQHE in the first excited LL, observed by Eisenstein 
{\sl et al.}\cite{exp3}

\section*{Acknowledgements}

We acknowledge fruitful discussions with M.\ P.\ A.\ Fisher, R.\ Mani,
R.\ Moessner, R.\ Morf, and A.\ W\'ojs. This work was supported by the Swiss 
National Foundation for Scientific Research under grant No.~620-62868.00.

\end{document}